\documentclass[prd, preprint, longbibliography, 11pt]{revtex4-1}

\usepackage{amsmath}
\usepackage{amssymb}
\usepackage{setspace}
\usepackage{graphicx}
\usepackage{natbib}
\usepackage{float}
\usepackage{tensor}
\usepackage[utf8]{inputenc}
\usepackage{amsfonts}
\usepackage{braket}
\usepackage{esint}
\usepackage{breqn}
\usepackage{IEEEtrantools}

\begin{document}
 
 %

\begin{center}
{ \large \bf The Pollen and the Electron: A Study in Randomness
 }


\vskip 0.1 in

{\large{\bf Priyanka Giri and Tejinder P.  Singh }}


{\it Istituto Nazionale di Fisica Nucleare, Pisa 56127, Italy}\\
{\it Tata Institute of Fundamental Research,}
{\it Homi Bhabha Road, Mumbai 400005, India}\\
 {\tt priyanka.giri@phd.unipi.it,  tpsingh@tifr.res.in}

\end{center}

\centerline{\bf ABSTRACT}
\noindent  The random motion of a pollen grain in a glass of water is only apparently so. It results from coarse-graining an underlying deterministic motion - that of the molecules of water colliding with the grain. Not observing degrees of freedom on smaller scales can make deterministic evolution appear indeterministic on larger scales. In this essay we attempt to make the case that quantum indeterminism arises in an analogous manner, from coarse-graining a deterministic (but non-unitary)  evolution at the Planck scale. The underlying evolution is described by the theory of trace dynamics, which is a deterministic matrix dynamics from which quantum theory and its indeterminism are emergent. One consequence of the theory is the Karolyhazy uncertainty relation, which implies a universal upper bound to the speed of computing, as noted also by other researchers.


\section{The pollen and the electron}
When you look at a pollen grain in a glass of water under a microscope, the grain exhibits random movement. This is the famous Brownian motion. You would be forgiven for thinking that this random motion of small particles suspended in a fluid is a law of nature. Something that is beyond the scope of Newtonian mechanics. But of course physicists have found out that this apparent randomness is a consequence of our ignorance. The molecules of water are colliding with  the pollen in accordance with Newton's laws, but because of inevitable statistical fluctuations in the number of molecules hitting the grain, the motion of the grain appears stochastic. Randomness is a consequence of coarse-graining, and of not examining the perfectly deterministic motion on molecular resolution scales. The quantitative derivation from atomic theory, of observed parameters of Brownian motion, fully supports this inference.

Now sample this. A beam of electrons is passing through a Stern-Gerlach apparatus, one at a time. Each electron has been carefully prepared to have its spin aligned  say forty-five degrees to the $+z$ direction. And we want to measure the spin of the electron along the $\pm z$ axis. This is what the experimentalist finds. Some electrons register their spin as $+z$ and some register it as $-z$. The outcome is unpredictable and random, but the outcomes are found to obey the Born probability distribution. Well, the electron was evolving according to a perfectly deterministic law (the Schrodinger equation, or more precisely,  the Dirac equation). Moreover the Stern-Gerlach apparatus is itself made of elementary particles which obey deterministic quantum mechanics. Where then does the randomness come from, and where do the probabilities come from? Quantum mechanics has no answer to this question. Could it be that, like the case of the pollen grain in water, the randomness is a consequence of our ignorance of some underlying microscopic dynamics? Or is randomness a fundamental property of the quantum measurement process, not to be questioned any further? Is the quantum mechanical description of nature like how we describe water as a thermodynamic fluid, and is there a deeper deterministic microscopic theory underlying QM, same way as atomic theory underlies the fluid that is water. Or is QM the exact ultimate dynamical law of nature? Physicists have struggled with this question ever since quantum theory was discovered.

At the heart of the conundrum is the quantum linear superposition principle, which asserts that a quantum system prepared in a superposition of two or more eigenstates of an observable stays in that superposition for an infinite time. Until and unless the quantum system meets a classical measuring apparatus, when superposition is broken, and the quantum system randomly `collapses' to one or the other of the superposed states. Is this randomness fundamental? Or is it a result  of coarse-graining an underlying deterministic theory - the analog of the molecular resolution of the fluid that is water. Here we would do well to remember that quantum superposition has been tested in the laboratory only upto objects made of about 25,000 elementary particles. Does superposition hold for objects larger than this? Maybe yes, maybe no. We don't know. But we know for sure that for macroscopic objects, made of say $10^{23}$ particles or more, such as chairs, tables, stars and galaxies, superposition does not hold. Even though a chair is  made of particles which by themselves obey superposition, the superposition vanishes when a large number of particles are bound together. Strange! And what classifies as large? $10^{10}$ particles, or $10^{16}$? We don't know!

In the 1980s, three Italian physicists, Ghirardi, Rimini and Weber, and an American, Phil Pearle, put forth a beautiful explanation \cite{Ghirardi:86,  Ghirardi2:90} to the above conundrum. They said, let us modify QM slightly. Instead of saying that a superposition of two states of a particle lasts forever, let us assume that it lasts for a very large, but {\it finite} time. Say for a time $T$ equal to the age of the universe: $T\sim 10^{17}$ s. After a mean life-time $T$ , the superposition spontaneously and randomly collapses to one of the many superposed states. This tiny change to QM is enough to take us home. For now if two particles were entangled together, the superposition would collapse if either one of them collapses, taking the other particle with it. So that the superposition life-time is now halved to $T/2$. If three particles are entangled then life-time is $T/3$, and so on. If $N$ particles are entangled, superposition life-time is down to $T/N$, and if $N$ is as large as $10^{27}$, like say in a chair, this life-time is as small as $10^{-10}$ s. Too small to be easily observable. Such an elegant solution to what happens in the Stern-Gerlach apparatus! Microscopic superpositions are very long-lived; but because of entanglement and spontaneous collapse, macroscopic superpositions are extremely short-lived.

This so-called theory of spontaneous localisation is currently being tested for, in laboratories around the world. Dynamical randomness has been introduced into what was earlier a deterministic quantum mechanics. The electron is behaving like the pollen in water - only, the random hits are extremely rare for an electron [hence superposition is long-lived]. But if the electron is replaced by a larger object such as a chair made of many, many particles, the hits become very very frequent [superposition is extremely short-lived]. But wait a minute. The pollen is being randomly hit by molecules of water. Who is doing the random hitting, when the electron meets the Stern-Gerlach measuring apparatus, and who is hitting the chair, to keep it classical?! The answer is profound; it takes us to the deepest reaches of space-time, the Planck scale, where lengths are as small as $10^{-33}$ cm, and times are as small as $10^{-43}$ s.

\section{Atoms of space-time-matter: predictability regained}
We well know that the gravitational effects of bodies are described by Newton's inverse square law of gravitation. Or, in the relativistic case, by Einstein's general theory of relativity. But these laws are for classical bodies. What is the gravitational effect of an electron, say when it is in a superposed state, having just passed through the two slits in a double slit interference experiment? How to describe gravity when quantum superpositions are present? 

A quantum particle in a superposition of states is delocalised; it is wavy and in a sense is everywhere. Its gravitational effect is also everywhere. There is then no meaning to distinguishing the particle from its gravitation. The source and the field become one and the same. And if the universe consisted entirely of such quantum particles [no classical bodies present] it is then no longer meaningful to talk of classical space-time. For, space is that which is between classical bodies, and time is that which is between classical events.

In such a situation, we talk of `atoms' of space-time-matter [STM] \cite{maithresh2019}. An electron together with its gravitation is an STM atom - the STM electron. It is described by a matrix, whose elements are Grassmann numbers. Grassmann numbers anti-commute with each other; the square of every Grassmann number is zero. Every STM matrix can be written as a sum of a bosonic matrix and a fermionic matrix. In a bosonic matrix, the elements are even-grade, being made of product of even number of Grassmann elements. The bosonic matrix describes the would-be-gravity part of the STM atom. The fermionic matrix has elements which are odd-grade Grassmann, and describes the would-be-matter part of the STM atom, e.g. the electron.

At the Planck scale, the dynamics of these matrix-valued STM atoms is a matrix dynamics \cite{Adler:04}. But this dynamics is not quantum theory. Nor is there any longer a classical space-time. Rather, these STM atoms live in a Hilbert space, endowed with an algebra which can be mapped to a non-commutative geometry [as in the programme of Alain Connes and collaborators \cite{Connes2000}]. Such a 
non-commutative algebra comes naturally equipped with a (reversible) time parameter, known as Connes time. The dynamics is simple to picture, as follows: Consider a classical mechanical system described by a set of configuration variables and canonical momenta, and a Lagrangian, from which the equations of motion arise. Now, raise each real-number valued dynamical variable to the status of a matrix (equivalently operator). The Lagrangian itself becomes a matrix valued  polynomial; take the matrix trace of this polynomial. Define this trace Lagrangian (a c-number) as the new Lagrangian, and its integral over Connes time as the new action. Also, raise each space-time point to the status of an operator - this is the essence of non-commutative geometry: the geometric degrees of freedom no longer commute with each other. A space-time operator together with the matrix-valued matter variable define the STM atom (a Grassmann matrix), and lead to a natural and elegant action principle. The dynamical degrees of freedom do not commute, and in fact obey arbitrary time-dependent commutation relations. The variation of the action with respect to the matrix variables gives equations of motion for the STM atom, which evolve in Hilbert space with respect to Connes time. Very significantly, the Hamiltonian of the STM atom is not self-adjoint; but has a tiny anti-self-adjoint part. A large collection of STM atoms, together with their dynamics, defines the fundamental universe \cite{Singh:sqg, singh2019qf, Singhdice}. {\it This dynamics is deterministic and
 time-reversible. Predictability is regained at the Planck scale}!

However, in the laboratory, we do not observe this Planck-scale matrix dynamics, wherein rapid variations occur over Planck time scales. Rather we observe a coarse-grained dynamics, at much lower energy scales; equivalently over coarse-grained time intervals much larger than Planck time. This averaging is in the same spirit in which averaging the molecular dynamics of many many molecules of water defines the thermodynamic properties of the fluid that is water. Except than, now the averaging is not over many STM atoms, but over the many Planck times that occur in defining one coarse-grained Connes time instant/interval. In other words, we want to know what is the mean motion of an STM atom, after the rapid Planck scale variations have been smoothed out. Beautifully so, as long as the anti-self-adjoint part of the STM Hamiltonian can be neglected, this mean dynamics is the same as that given by quantum theory! There is still no space-time, but the commutation relations satisfied by the averaged matrix variables are now those of quantum theory. The averaged STM atom is akin to the pollen, and the underlying rapid variations on the Planck scale are akin to the molecules of water which push the pollen around. Under certain circumstances, which we now describe, these rapid variations become significant and disrupt the mean motion. These rapid variations provide the sought for random hits, negligible for a single STM atom, but crucial when many STM atoms get entangled.

When a sufficiently large number of STM atoms get entangled, an effective length scale associated with the entangled system goes below Planck length, the imaginary part of the Hamiltonian becomes significant, and the coarse-graining approximation leading to emergent quantum theory breaks down. The individual sub-Planck motion of each STM atom tries to pull the mean-field entangled system its own way. These pulls are inevitably random, because there are so many STM atoms in the entanglement. These random pulls are the equivalent of the molecules of water that push the pollen around. These extremely frequent random hits result in spontaneous localisation and classicality of the fermionic (matter) part, accompanied by the emergence of a classical space-time geometry obeying the laws of classical general relativity. When an electron passes though the Stern-Gerlach apparatus, its exact time of arrival at the apparatus, down to Planck scale resolution, is crucial. This arrival time decides which STM atom's hit (from the apparatus) comes into play, and determines which state the electron will evolve to. The randomness of the time of arrival of the electron makes a perfectly deterministic (though non-local) underlying dynamics appear random. The electron meeting a measuring apparatus is precisely like the pollen in a glass of water, in so far as determinism and randomness are concerned. But the Planck scale space-time-matter foam looks extremely different from the classical space-time and quantum matter fields we are accustomed to.

One far-reaching consequence of the STM matrix dynamics is that it predicts space-time to be holographic. The theory predicts \cite{SinghqgV2019, Singh:DE} that if one were to use a measuring apparatus to measure a length $L$, there will always be a minimum uncertainty $\Delta L$ in this measurement, given by
\begin{equation}
(\Delta L)^3 \sim L_P^2 \; L
\label{kul}
\end{equation}
where $L_P$ is Planck length. This defines the smallest possible fundamental volume inside of a spatial region of size $L^3$, implying that the number of information units grows as $L^3 / (\Delta L)^3 \; \sim \; (L/L_P)^2$. This indeed is the holographic principle, i.e. the amount of information in a region increases, not as its volume, but as its area. This same principle also leads us to a derivation of the Bekenstein-Hawking entropy of a black hole (one-fourth the area of the black hole) from the microstates of the STM atoms which constitute the black hole \cite{maithresh2019b}. This holographic inference also implies an upper bound on the ability of a computer to compute! Furthermore, the spontaneous localisation of a sufficiently large number of entangled STM atoms necessarily results in the formation of a black hole - that simplest and most beautiful of classical objects, characterised only by their mass, charge, and angular momentum. And if one tries to make a probing device which can go sub-Planckian and experience the deterministic matrix dynamics at play there, the device necessarily becomes a black hole. It cannot communicate the knowledge of deterministic dynamics to the outside universe. It is as if the STM atom has two pristine states: one being the matter-dominated state (i.e. the quantum electron), and the other being the gravity dominated state (the black hole). The quantum electron and the classical black-hole are dual states of each other - the  former is the ultimate particle, and the latter is the ultimate computer.

\section{Limits to computability: black hole as the ultimate low-energy computer}
Analogous to the uncertainty relation for lengths, the STM matrix dynamics also predicts an uncertainty relation for measurements of time. If a device is used to measure a time interval $T$, there will be a minimum resolution / uncertainty $\Delta t$, given by
\begin{equation}
(\Delta t)^3 \sim t_P^2\; T
\label{klt}
\end{equation}
where $t_P$ is Planck time, defined by $t_P^2 = G\hbar /c^5$, and numerically equal to $10^{-43}$ s. This lower limit sets a bound on the speed of a computer. No computer can ever complete one computational step in a time less than $\Delta t$. If the computer runs for a time $T$, the memory space $K$ available for computation is of the order $K\sim T/\Delta t \sim (\Delta t/t_P)^2$. This implies that $K/(\Delta t)^2 \sim t_P^{-2} \sim 10^{86}$ s$^{-2}$ is a universal constant. One could try to increase a computer's computing power by making it run longer (higher $T$), but that reduces its computing speed $\nu\equiv 1/\Delta t$, hence this universal bound. No computer can be so long-lasting and so efficient as to beat this bound on $K\nu^2$. For comparison, our laptops perform $10^{10}$ operations / s$^{2}$. This universal bound has also been derived by other researchers earlier, using semiclassical heuristic arguments \cite{Ng, Lloyd}. However, ours is the first rigorous derivation stemming from quantum foundations and quantum gravity. There is an upper bound on computability because a chair cannot be in more than one place at the same time!

The Bekenstein-Hawking entropy of a black hole is far, far higher than the Boltzmann entropy for normal systems of comparable mass. This entropy can be equated to the Shannon entropy one associates with information, because the entropy comes from coarse-graining over the microstates of STM atoms. Since a high entropy means a high amount of information hidden away, a black hole is the most efficient information storage device. The number of bits  being given by $(L/L_P)^2$, as we saw above, where $L$ is the linear extent of the black hole. If $L$ is one centimetre, this gives $10^{66}$ storage bits per cubic cm. This would make a black hole as the ultimate computer. Applying the above uncertainty relation to a black hole, and taking $L\sim GM/c^2$ to be the size of the black hole, where $M$ is the mass of the black hole, and assuming that computational speed is determined by  $c\Delta t = L$, we get that the life-time of the black hole computer is $T\sim G^2 m^3/\hbar c^4 = t_P (m/m_P)^3$. Here $m_P$ is Planck mass. Remarkably, this also happens to be the time-scale over which a black hole disappears as a result of Hawking evaporation! This consistency of arguments establishes a fundamental connection between quantum unpredictability and limits to computability. It is easy to check that the black hole evaporation time satisfies the computability bound $T\nu^2 \sim t_P^{-2}$. Thus a computer made with the same qualities as a black hole would be the ultimate low-energy computer.

In order to treat the  black hole as a computer,  it has to pass the Turing completeness test: is the system  a Turing machine or not? The system should be able to simulate the Turing machine irrespective of runtime and memory use. Black holes too can act as a Turing machine under certain limitations like any other physical computer. The tape for the Turing machine is the black hole’s contents itself. And, the requirement for inextensible tape can be achieved by increasing the mass of the black hole. The external observer will move in order to shift the tape and read the output via Hawking radiation. 
 Finally, there exists a set of instructions which form a Turing-complete language from these physical components: the state of a position on the tape may be changed by irradiating that particle with light, the head may be moved, and information may be read by the head from the Hawking radiation \cite{Andrews}. 
 
 The above considerations  are at low energies, outside the black hole, and hence approximate. The picture changes dramatically
 if we enter the black hole or approach Planck energies, where space-time is lost. 
 
 \section{Predictable quantum computing at the Planck scale}
 At the Planck scale, we have a deterministic matrix dynamics of atoms of space-time-matter. These evolve with respect to Connes time $\tau$. If ever a computer were to be made at the Planck scale, these STM atoms would be the entities to make it from. It would be a quantum computer alright, and still a Turing machine, but with a difference. Recall that the matrix dynamics, though deterministic, is non-unitary. And there is no classical space-time. So we can make a quantum computer by superposing states of one or more STM atoms, but this superposition will not last forever. Depending on how many STM atoms are entangled in the superposition, there will be a deterministic decay to one of the states, sooner or later, according to Connes time. The beauty of this quantum computer is that when a measurement is made, the outcome is predictable, not random nor probabilistic. The predictable nature of such outcomes gets rid of errors that might be associated with the stochastic nature of the outcome (collapse of the wave function) in a conventional quantum computer. And it is decidedly advantageous to have a completely predictable quantum computer, rather than one whose final outcomes are unpredictable because of the notorious quantum measurement problem.
 
 How might one realise such a Planck scale computer? One way is to send the observer inside the black hole, all the way to the classical black hole space-time singularity. In our matrix dynamics there is no such singularity though,  it having been replaced by the finite dynamics of STM atoms. But our poor observer would nonetheless be crushed to oblivion in these hostile environs. We may instead conjure up a Maxwell's demon, who watches and manipulates these STM atoms, treats their matrix dynamics as an initial value problem, and quantum computes with them. The predictable nature of the outcomes makes this a kind of hyper-computation, going beyond the reach of classical and quantum computers as we understand them at present. Sadly though, no such demon can communicate the results of such computations to his human friends outside the black hole. And the reason is illuminating!
 
 While at it, we point out that in our theory there is no such thing as the black hole information loss paradox. The conventional statement of the paradox is that an initial quantum state of a matter field in the vicinity of the black hole, evolves unitarily according to quantum theory, and gives rise to  thermal Hawking radiation. Complete evaporation of the black hole would convert the initially pure quantum state into the mixed state that Hawking radiation is, thus violating unitarity of quantum theory. For us though, this is not how it works. Let us treat as a full system the quantum matter field and the {\it classical} black hole. The very process of black hole formation is non-unitary, it having resulted from the spontaneous localisation of an enormous collection of entangled STM atoms. And we know that spontaneous localisation results from non-unitary evolution: the associated length scale to which localisation takes place is given by $L_S = L_{eff}^3 /L_P^2$. Here, $L_{eff}=L/N$ is the effective Compton wavelength of an object made of $N$ entangled particles, each having a Compton wavelength $L$. If $L_{eff}$ is less than Planck length, spontaneous localisation results in a black hole, and this requires that the object should at least be as massive as Planck mass. Now, this process, though non-unitary, is deterministic. The information about black hole formation is coded in the anti-self-adjoint part of the fermionic Hamiltonian of entangled STM atoms. Upon Hawking evaporation, this information is not lost. It is present in the evaporated radiation, but at sub-Planckian length scales. To detect this correlation of entanglements in the Hawking radiation one will have to probe the radiation at sub-Planck scales, but that will again result in the probe becoming another new black hole!

Nonetheless, in spite of black holes turning up all over the place, it is possible to make a fully predictable quantum computer using our matrix dynamics, in the laboratory, at least in principle. The computer and the apparatus that measures the outcome are the two sub-systems of a combined deterministic system with the condition that the total mass is less than Planck mass. So that at all events a black hole formation is avoided. To make the quantum computer, a set of entangled STM atoms is employed, with  the initial conditions of the matrix dynamics [i.e. initial values of matrix components] precisely known. Then the quantum computation part proceeds just as in a conventional quantum computer, noting that the number of qubits is small enough that the spontaneous localisation lifetime is much longer than the duration of the computation. When the time comes to measure the output (which will be one of the matrix components), the system interacts (deterministically) with a much larger collection of STM atoms (a second quantum system). This step is analogous to an electron arriving at the photographic plate (measuring apparatus)  in a double slit interference experiment. Except that, now the plate is replaced by a large entangled quantum system with total mass such that the non-unitary evolution becomes significant, and spontaneous localisation sets in rapidly, on a measurable time scale. The quantum superposition present in the quantum computer will decay, deterministically, to  a predictable outcome, which can be programmed algorithmically, knowing the rules of the matrix dynamics. We have a quantum Turing machine with predictability.

Predictability, or the lack of it, and computability, or the lack of it, are not absolute givens. These properties are determined by the physical laws of nature. We have shown that fundamentally, nature is deterministic and predictable. Einstein was right about this, even though he was wrong in hoping that the physical world is local. Quantum unpredictability is only a consequence of our ignorance of the world at Planck scale, much the same way as to why the random motion of a pollen grain in a glass of water is only apparently unpredictable. These new developments impact on how we think about computability, and have implications for the future of computers. Who knows, future developments in physics might lead to re-thinking on undecidability and uncomputability as well? Mathematical theorems are based on axioms. These axioms often tend to reflect our perceptions of the physical world, and these latter change with our understanding of physical theories.


\bigskip

\centerline{\bf REFERENCES}

\bibliography{biblioqmtstorsion}

\end{document}